\documentclass[11pt]{article}

\usepackage{epsfig}
\newcommand{\be}{\begin{equation}}
\newcommand{\ee}{\end{equation}}

\newcommand{\lfrac}[2]{\mbox{${#1}\over{#2}$}}

\begin{document}
\title{Coherent States}
\author{Peter W. Milonni and Michael Martin Nieto \\
Los Alamos National Laboratory}
\date{}
\maketitle

{\bf Coherent states} (of the harmonic oscillator) were introduced by Erwin Schr\"odinger (1887-1961) at the very beginning of quantum mechanics in response to a complaint by Lorentz that Schr\"odinger's 
wave functions did not display classical motion. Schr\"odinger obtained solutions that were Gaussians having the width of the ground state. The expectation values of the coordinate and momentum for these Gaussian solutions oscillate in time in just the same way as the coordinate and momentum in the classical theory of the harmonic oscillator. 

In modern parlance Schr\"odinger's solutions are 
the 2-parameter ($\langle x\rangle,\langle p\rangle$) states
\begin{equation}
\psi_\mathrm{cs} = [2\pi (\Delta x)^2]^{-1/4} \exp
\left[-\left(\frac{x-\langle x\rangle}{2\Delta x}\right)^2+i\frac{\langle p\rangle x}{\hbar}\right]
\end{equation}
satisfying equality in the uncertainty relation 
\begin{equation}
(\Delta x)^2 (\Delta p)^2 \ge {\hbar^2 \over 4}
\end{equation}
and having ``widths" equal to those of the ground state, $(\sqrt{2}\Delta x)\equiv (\hbar/m\omega)^{1/2}$.\footnote{Squeezed states, whose width oscillates with time, were introduced in 1927 by E. H. Kennard.  They are a 3-parameter set of Gaussians whose widths are {\it not} that of the ground state.}  These can be called {\it minimum uncertainty coherent states}.

In the 1960's there was a reawakening of interest in these states in terms of the boson operator formalism.  Two other, equivalent formulations of coherent states were obtained.  The first yields the {\it annihilation operator coherent states}, $|\alpha\rangle$, defined by
\begin{equation}
a|\alpha\rangle = \alpha |\alpha\rangle,
\end{equation}
where $a$ ($a^{\dag}$) is the annihilation (creation) operator.  
The second yields the {\it displacement operator coherent states} 
\begin{equation}
|\alpha\rangle \equiv
	D(\alpha)|0\rangle  = \exp[\alpha a^{\dagger} - \alpha^* a] |0\rangle .	
\end{equation}
The real and imaginary parts of the complex number $\alpha$ are the two parameters which give the solution as 
\begin{equation}
|\alpha\rangle = \exp\left[-\frac{1}{2}|\alpha|^2\right] \sum_{n=0}^\infty \frac{\alpha ^n}{\sqrt{n!}} 
|n\rangle , 
\end{equation}
where $|n\rangle$ are the number states, i.e., the energy eigenstates of the harmonic oscillator. From the Hermite polynomial generating function these can be shown to be identical to the Gaussians of the minimum-uncertainty coherent states, where 
\begin{equation}
\mathrm{Re} ~\alpha = \langle x\rangle \left(m\omega \over 2 \hbar\right)^{1/2}, ~~~~~~~~~~ \mathrm{Im} ~\alpha = \langle p\rangle\left(1 \over 2 m\omega  \hbar\right)^{1/2}.
\end{equation}

These ideas have been applied to non-harmonic systems, involving different symmetries and/or potentials.  There the coherence properties are not as strong in general, since it is the equally-spaced levels of the harmonic oscillator which allow the system never to decohere if there is no damping or excitation.


An especially interesting system is the even- and odd-coherent states 
(``cat" states).  They are higher-power states, eigenvalues of $aa$. They are given by [$(\hbar,m)\equiv 1$]
\begin{eqnarray}
|\alpha;+\rangle &=& [\cosh{|\alpha|^2}]^{-1/2}
      \sum_{n=0}^{\infty}\frac{\alpha^{2n}}{\sqrt{(2n)!}}|2n\rangle 
\rightarrow  \psi_{+}(x),  \\
|\alpha;-\rangle &=& [\sinh{|\alpha|^2}]^{-1/2}
\sum_{n=0}^{\infty}\frac{\alpha^{2n+1}}{\sqrt{(2n+1)!}}|2n+1\rangle 
\rightarrow  \psi_{-}(x).
\end{eqnarray}
\begin{equation}
\psi_{\pm}(x)
    = \frac{e^{-i2 x_0 p_0}\left[\exp[-\lfrac{1}{2}(x-x_0)^2] e^{i p_0 x}
	    \pm \exp[-\lfrac{1}{2}(x+x_0)^2]e^{ - i p_0 x}\right]}
      {2^{1/2} \pi^{1/4}\left[1 \pm \exp[-(x_0^2+p_0^2)]\right]^{1/2}}.
\label{eocs}
\end{equation}


\begin{figure}[h!]
\begin{center}
\psfig{figure=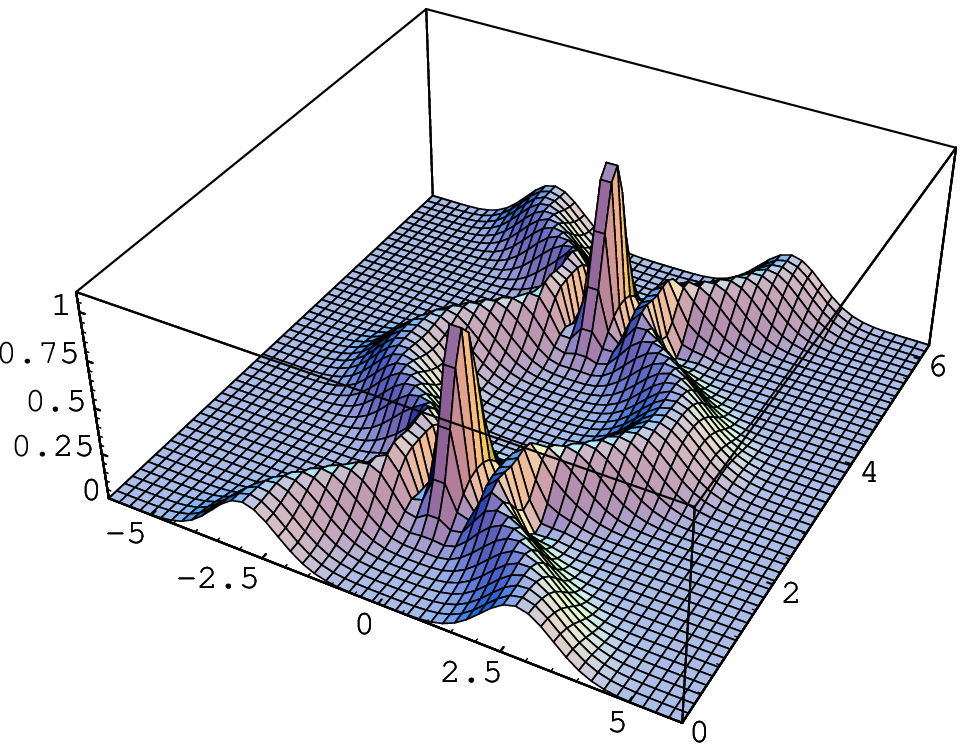,width=2.75in}
\hskip 10pt
\psfig{figure=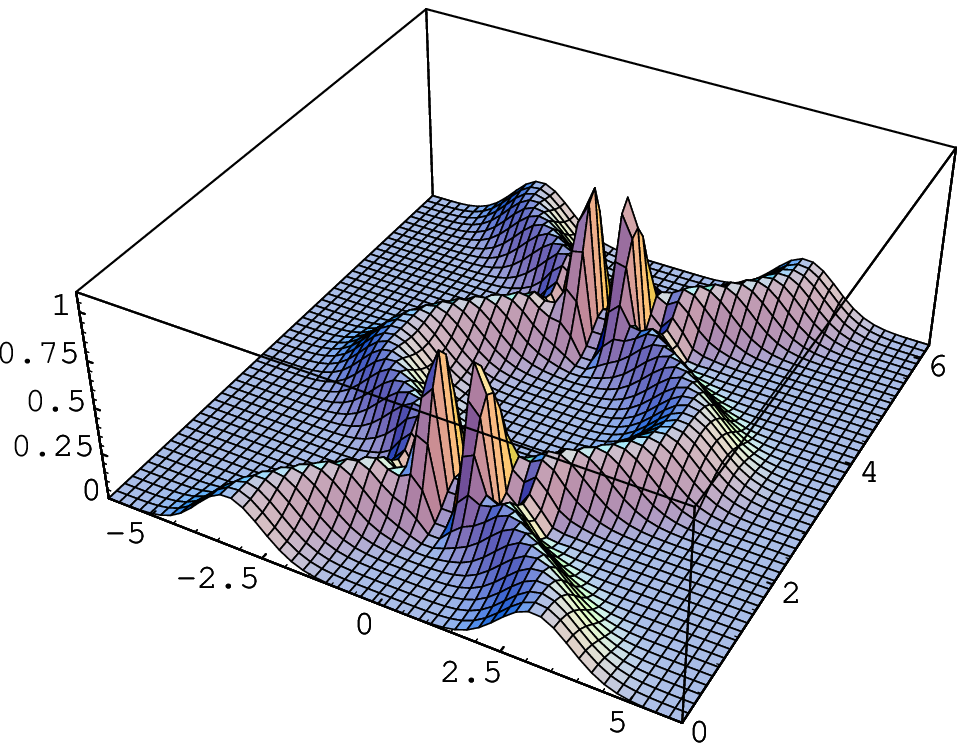,width=2.75in}
  \caption{The time evolution of 
the even- and odd-coherent states $\rho_{\pm}(x,t)$ 
The initial conditions are $x_0 = 2^{3/2}$ and $p_0 = 0$.
The position is along the $x$-axix, time is along the $y$ axis, 
and the $Z$-axis displays the probability density.
\label{fig:eocsfig1}}
 \end{center}
\end{figure}

The wave packets of these  states are two Gaussians, at positions $\pi$ 
apart in the phase-space circle. The Gaussians keep their shapes as they move as a normnal coherent state would in time evolution, until they overlap.  When the even states, composed of $n=0,2,4,\dots$ number states. interfere, they have a maximum central peak. (See the left graph in Figure 
\ref{fig:eocsfig1}.) 
The odd states are composed of 
$n=1,3,5,\dots$ number states.   When the odd Gaussians interfere there is
a central minimum and two slightly smaller peaks on each side. (See the right graph in Figure \ref{fig:eocsfig1}.)  

These states have been observed experimentally (Monroe et al.).


The coherent states have been especially useful in quantum optics. Each mode of the electromagnetic field may be described
formally as a harmonic oscillator, and different quantum states of the oscillator correspond to different states
of the field. The field from a single-mode laser operating far enough above threshold can be described for many purposes as a coherent state; it differs from a coherent state in that its phase drifts randomly. But its photon counting statistics  and other properties make the light from a single-mode laser practically indistinguishable from a coherent state.

The quantum theory of optical coherence is based on ``normally ordered" products of lowering and raising
operators $a$ and $a^{\dag}$ which act respectively as photon annihilation and creation operators. The fact that coherent states are eigenstates of lowering operators implies that the expectation value of a normally ordered field operator product $f(a,a^{\dag})$ reduces to the deterministic function$f(\alpha,\alpha^*)$ for a coherent state.  A coherent state of the field therefore comes closest to the idealized classical stable wave in which there are no random field fluctuations. Thus a coherent-state field exhibits maximal fringe visibility or ``coherence" in a Michelson interferometer, for instance, and it is maximally coherent as well when more complicated interference effects involving higher orders of field products are considered. 
\\ \\
{\bf Literature}\\
{\it Primary} \\
$\bullet$ \  \ Schr\"odinger, E.: `Der stetige \"Ubergang von der Mikro- zur Makromechanik', {\it Naturwiss.} {\bf 14}, 664 (1926). Translated into English in Schr\"odinger, E.: 
{\it Collected Papers in Wave Mechanics}.  London, Blackie\& Son (1928), p. 41\\
$\bullet$ \ \ Glauber, R. J.: `Coherent and Incoherent States of the Radiation Field', {\it Physical Review} {\bf 131}, 2766 (1963)\\
$\bullet$ \ \ Klauder, J. R.:  `Continuous-Representation Theory',  {\it J. Math. Phys.} {\bf 4}, 1055, 1058 (1963)  \\
$\bullet$ \ \ Sudarshan, E. C. G.:  `Equivalence of Semiclassical and Quantum Mechanical Descriptions of Statistical Light Beams', 
{\it Phys. Rev. Lett.} {\bf 10}, 277 (1963) \\
{\it Secondary}\\
$\bullet$ \ \ Klauder, J. R., B.-S. Skagerstam: {\it Coherent States -- Applications in Physics
and Mathematical Physics}. Singapore, World Scientific (1985)\\
$\bullet$ \ \ Mandel, L., E. Wolf, {\it Optical Coherence and Quantum Optics}. Cambridge: Cambridge University Press (1995)\\
$\bullet$ \ \ Nieto, M. M., D. R. Truax: 
`Higher-Power Coherent and Squeezed States,' {\it Optics Comm.}
{\bf 179}, 197 (2000) \\
$\bullet$ \ \ Monroe, C, D. M. Meekhof, B. E. King, and D. J. Wineland: 
`A ``Schrodinger Cat" Superposition State of an Atom,' {\it Science} {\bf 272}, 1131 (1996)

\end{document}